# Assessing the economic benefits of space weather mitigation investment decisions: Evidence from Aotearoa New Zealand

Edward J. Oughton[1*], Andrew Renton[2], Daniel H. Mac Marnus[3], Dennies Bor[1], Craig J. Rodger[3]

[1]Geography and Geoinformation Sciences, George Mason University, Fairfax, Virginia, USA. [2]Transpower New Zealand Ltd, Wellington, New Zealand. [3]Department of Physics, University of Otago, Dunedin, New Zealand *Corresponding author (eoughton@gmu.edu)

## Abstract

Space weather events pose a growing threat to modern economies, yet their macroeconomic consequences remain underexplored. This study presents the *first dedicated economic assessment* of geomagnetic storm impacts on Aotearoa New Zealand, quantifying potential gross domestic product (GDP) losses across seven conservative disruption and mitigation scenarios due to an extreme coronal mass ejection (CME). The primary focus is on the damaging impacts of geomagnetically induced currents (GICs) on the electrical power transmission network. We support space weather mitigation investment decisions by providing a first-order approximation of their potential economic benefits, using best-in-class scientific models, via a coupled physics-engineering-economic spatial modelling framework. Recognising uncertainty in the economic interpretation of power outage impacts, we compare four different estimation methods. In the most severe unmitigated scenario, estimated GDP losses reach NZ$3.58 billion (0.98% of annual GDP). Targeted GIC-informed scenarios still produce material losses, with no mitigation reaching up to NZ$1.48 billion (0.41% of annual GDP). Mitigation substantially reduces these impacts. Operational strategies, including optimized switching and islanding, achieve benefit-cost ratios as high as 330 to 1, while physical protections such as GIC blocking devices produce returns up to 34.4 to 1. When also acknowledging additional unmodelled impacts, including multi-billion losses in capital equipment and long-term revenue, the economic rationale for pre-emptive mitigation becomes even more pertinent.

## Plain language summary

Space weather events can disrupt electricity systems and cause widespread economic damage. Despite this risk, their broader economic impacts have not been well studied. This research provides the first evaluation of how an extreme solar storm could affect New Zealand's economy by estimating how power disruptions could reduce economic activity under several realistic scenarios. A severe but plausible solar storm without mitigation could cause up to NZ$3.58 billion in lost economic activity, equivalent to almost 1% of New Zealand's annual economy. Targeted exposure scenarios could still lead to losses of up to NZ$1.48 billion. Importantly, relatively inexpensive mitigation measures can deliver large economic benefits. For example, operational strategies such as grid switching and temporarily separating parts of the network provide strong benefits, with up to NZ$330 in avoided economic losses for every NZ$1 invested. Physical protections, like devices that block harmful currents, also offer strong returns on investment.

## Key points

- A plausible no-mitigation solar storm could cost NZ$0.74-3.58 billion in lost GDP (0.2-0.98% of annual GDP).
- Mitigation measures achieve benefit-cost ratios as high as 330 to 1.
- Investment in space weather mitigation is highly cost-effective.

## 1. Introduction

Space weather refers to a variety of solar-driven phenomena, including coronal mass ejections (CMEs), solar flares, and solar particle events, which can disturb Earth's magnetosphere and upper atmosphere (Hapgood *et al.*, 2021; Buzulukova and Tsurutani, 2022). An extreme CME can induce geomagnetically induced currents (GICs) in extra high voltage (EHV) electricity transmission networks (>200 kV), posing a risk to the stability of the grid. Importantly, this critical infrastructure underpins technology-dependent economies across the world. In recent decades, researchers have increasingly sought to quantify the economic impacts of major natural disasters on critical infrastructure such as earthquakes, hurricanes and space weather events, motivating the assessment here.

The most intense geomagnetic storm on record is the Carrington Event of 1859 (Love *et al.*, 2024; Thomas *et al.*, 2024). While there are no direct economic loss figures from 1859, the event serves as a benchmark for worst-case scenarios given the estimated size and duration. Such an extreme event would impact multiple countries and likely induce cascading failures across sectors, underscoring why Carrington-class storms frame the upper bound of space weather risk.

A more recent example is the geomagnetic storm of March 13, 1989, which caused the Hydro-Québec power grid in Canada to collapse. GICs saturated transformers and destabilized the grid, resulting in a province-wide blackout that lasted about 9 hours and affected 6 million customers (Allen *et al.*, 1989; Bolduc *et al.*, 1998). The direct physical damage from the 1989 storm was relatively limited, for instance with a report by Hydro-Québec noting about $6.5 million in equipment damage from overvoltages with the net cost of the Quebec blackout (including utility losses and immediate economic impacts) estimated at $13.2 million (in 1989 USD) (Bolduc, 2002, p. 20). A US assessment noted that for a similar event, a space weather induced blackout in the US Northeast could cost on the order of $3–6 billion (1990 USD) (Barnes and Dyke, 1990). This wide range of figures highlights an important point, that economic losses escalate dramatically when indirect impacts on businesses and consumers are included, beyond immediate utility damages. The 1989 Quebec blackout, though a moderate event by space weather standards, revealed the large scale of losses possible from geomagnetic storms in modern grids.

In late October 2003, a series of intense solar storms (the "Halloween storms") struck Earth, causing global space weather disturbances. While North America avoided a major grid collapse, notable effects were recorded elsewhere. On 30th October 2003, GICs caused a power blackout in Malmö, Sweden, that lasted about an hour and cut power to ~50,000 customers (Pulkkinen *et al.*, 2005). The economic impact of this brief outage was limited (on the order of a few million dollars at most), but it underscores that even mid-latitude locations are not immune (30-60°). During the same event, transformers failed in South Africa surprising engineers who had assumed even lower latitudes were safe (Gaunt and Coetzee, 2007). Replacing large transformers is costly (often tens of millions of dollars each) and can take weeks if there is a spare, or years if a new replacement is sought, even under business-as-usual conditions, implying substantial economic repercussions. Compared to other hazards, there are still very few *realistic assessments* of space weather impacts in the peer-reviewed literature (Eastwood *et al.*, 2017).

In terms of recent events, New Zealand provides an interesting case for the May 2024 "Gannon" event given that it is one of the few examples where an operational national strategy was created, planned, and successfully followed in response to an evolving space weather event. Here, the national grid operator (Transpower) deployed a switching strategy to deal with field changes between 287-374 nT/min, producing recorded GIC currents peaking at 113 A compared to calculated values of 103 A. While this was not a Carrington-sized event, if the national switching plan had not been implemented it was calculated that peak values would have reached values of at least 200 A, illustrating how risk can be reduced via a thoroughly researched strategy. Given this context, it is subsequently very important to estimate both the potential economic impacts of such events, and the benefits of different potential investments in mitigation options. This is because investment in space weather mitigation is often a government-level policy decision based on socio-economic impact and therefore, peer-reviewed analysis of such impacts is critical for the formation of evidence-based policy.

Hence, the following two research questions are posited for investigation.

a) What might be the potential economic consequences of a major space weather event in Aotearoa New Zealand?
b) Which mitigation investment strategies are most effective, when considering both the costs of investment, and the benefits of avoiding lost Gross Domestic Product (GDP)?

As economic impacts from infrastructure disruption are sensitive to modelling assumptions, this paper reports results across alternative economic impact estimation methods rather than relying on a single macroeconomic approach. This allows the analysis to focus on the robustness of the mitigation investment case under methodological uncertainty.

Having now outlined the motivation for this research activity, the following section undertakes a literature review. A method is then presented in Section 3. Results are reported in Section 4, before being discussed in Section 5. Finally, conclusions are provided in Section 6.

## 2. Literature review

This review summarizes the current state of peer-reviewed literature on space weather's economic consequences, including modelling studies that estimate the potential costs of extreme events. Emphasis is placed on methodologies for economic impact assessment, especially for Input–Output (IO) models. The goal is to provide a comprehensive, global perspective on the economic risks posed by space weather, specifically because these events occur globally, with the review considering impacts around the world. While extreme geomagnetic storms are infrequent, even moderate storms have caused notable disruptions.

### 2.1. Extreme event scenarios and global impact assessments

While historical storms like 1989 and 2003 provide real data points, there is still concern whether we may experience the potential impact of a truly severe event, such as a modern-day Carrington Event. As we have not experienced this during the space age, researchers use simulations and scenario analysis to gauge the possible impacts on technology, infrastructure and the subsequent socio-economic implications. Several peer-reviewed studies have approached this from national and global perspectives.

One assessment modelled the impacts of widespread US power grid failure from an extreme storm using an IO economic framework (Oughton *et al.*, 2017). The most severe scenario envisioned a power outage covering the northern two-thirds of the United States (≈66% of the population). The results indicated a domestic economic loss of about US$41.5 billion per day of outage, plus $7 billion per day in international supply chain losses due to domestic firms being unable to produce goods for the international market. In other words, each day of nationwide blackout could cost nearly US$50 billion in GDP, with roughly half the losses accruing outside the immediate blackout zone via economic linkages. Even more moderate large-scale outage scenarios had sobering costs, as a storm leaving 23% of the US population without power sees daily US GDP loss estimated at US$16.5 billion, with an additional US$2.2 billion globally. These figures illustrate how indirect impacts via supply chain disruption can lead to large losses accruing, with the study finding that on average only ~49% of total losses were the direct result of power being out, while ~51% were indirect effects cascading through supply chains. Notably, the manufacturing

sector would bear the largest losses, and trading partners like China, Canada, and Mexico would also suffer billions in secondary economic damage when US industries go offline.

Another evaluation undertook a global economic modelling approach to appraise severe space weather impacts on interconnected economies via a global supply chain shock analysis (Schulte in den Bäumen *et al.*, 2014). Using a Multi-Region IO (MRIO) model, they simulated scenarios where power grid capacity is reduced by about 10% in key regions (North America, Europe, East Asia) for a period, akin to a "Quebec 1989" level event affecting those economies. They found that the global GDP could be reduced by ~3.9% to 5.6% (approximately US$2.4–3.4 trillion) over the year following such an event. Even if a storm's direct impacts were concentrated in, say, the US and Europe, the whole global economy could experience a non-trivial recessionary hit due to lost trade and productivity. There is debate as to whether the restoration period modelled here is realistic.

The insurance sector has also produced scenario analyses that, while not always peer-reviewed, are widely cited in the academic literature. The Lloyd's 2013 report (developed with Atmospheric and Environmental Research) posited an extreme geomagnetic storm scenario impacting the US East Coast. As mentioned, it projected US$0.6–2.6 trillion in economic costs and long-duration outages (Lloyd's of London, 2013). There are also concerns as to whether this assessment is plausible and realistic. For example, in another analysis by PwC in 2016 for the European Space Agency it was suggested that a 3-day blackout affecting major European cities would cost on the order of €5.7 billion (PwC, 2016). These estimates, while more modest, are seen to be more plausible and realistic, because operators would either actively choose to move assets offline to protect them, or protective equipment would do this automatically, prior to assets being catastrophically damaged. It would then be a case of needing to restart the grid in a matter of days, rather than power outages lasting up to multiple years (Oughton *et al.*, 2019). The difference has huge economic implications, as a days-long outage might be managed with limited economic harm, whereas a months-long grid recovery would be economically devastating. Consequently, scenario studies often explore ranges of severity to account for unknown outcomes (Eastwood *et al.*, 2018).

### 2.2. Methods for assessing the economic impact of space weather

Accurately assessing the economic impact of space weather events is challenging, requiring interdisciplinary methods. Unlike common natural disasters (e.g., hurricanes) where we directly observe property damage, space weather's effects are often indirect (e.g., lost electricity). Researchers have adapted several methodologies to quantify both direct and indirect losses (Oughton, 2018). Below we outline these key methods used in space weather studies, many of which draw on analogies from the literature on modelling natural hazard impacts.

One frequent approach is to estimate the direct losses due to service interruptions using the value of lost load (VoLL). For power outages, this often assigns a dollar value per unit of electricity not delivered. For example, the interruption cost to electricity consumers might be taken as US$5,000–15,000 per MWh of unserved energy (Abt Associates, 2017) (or ~NZ$20,000 per MWh in Aotearoa New Zealand). Multiplying by the estimated energy unsupplied during a blackout gives a rough cost. VoLL-based estimates are transparent and easy to compute, but they capture only the immediate utility-customer impact (analogous to direct business losses) and can vary widely by region and customer type. They do not account for broader supply chain disruptions.

Alternatively, IO models are a workhorse for disaster impact analysis and have been embraced in space weather studies. An IO model represents the interdependencies of industries in an economy through a matrix of purchases and sales. If a shock (like a power outage) reduces output in one sector, an IO model can estimate how upstream suppliers and downstream customers are affected in turn (Oughton *et al.*, 2017).

Similarly, MRIO approaches can model the propagation of impacts internationally (Schulte in den Bäumen *et al.*, 2014). Within the IO family, Leontief models are commonly used to estimate how reductions in final demand propagate upstream through supplier industries, whereas Ghosh models are used to examine how supply-side disruptions propagate downstream through purchasing industries. Thus, the Ghosh formulation is particularly relevant when the initial disruption is interpreted as a reduction in productive capacity, as may occur when sectors dependent on electricity are unable to operate during an outage. Generally, IO modelling is well suited to short-term scenarios, such as a sudden outage that reduces output in proportion to downtime. However, IO approaches can overestimate losses if used over longer periods, because they generally assume limited substitution and adaptation. In addition, supply-side IO specifications, including the Ghosh model, have been criticised when interpreted as behavioural quantity models for market economies (Oosterhaven, 2012, 2022). These critiques motivate the use of multiple economic impact methods in the present study, rather than relying on a single IO closure (Sprintson and Oughton, 2025). Despite this, IO analyses are valued for providing an upper-bound estimate of economy-wide impact and highlighting critical inter-sector linkages. Notably, risk analysts have applied IO models to other infrastructure failure scenarios (like the 2003 Northeast US blackout) with comparable findings that ripple effects can rival direct losses (Anderson and Geckil, 2003; Galbusera and Giannopoulos, 2018).

A more sophisticated approach leverages CGE models, which simulate the entire economy (producers, consumers, prices, and markets) and allow for behavioural adjustments. CGE models can account for businesses finding alternative inputs, consumers shifting spending, and price changes after a disaster, and therefore is more beneficial when modelling long-term power outages. In the context of space weather, CGE modelling has been less common so far, but analogous natural disaster research suggests it can yield valuable insights. For example, one CGE study examined a hypothetical 2-week power outage in California's Bay Area and found businesses would partially compensate by shifting production, reducing the total GDP loss compared to an IO estimate (Sue Wing and Rose, 2020). If applied to space weather scenarios, a CGE model might predict smaller losses than an IO model (thanks to economic resilience), but requires detailed data on how sectors adapt to power loss. As space weather impact assessment matures, CGE methods could complement IO results to provide ranges (with IO as a higher bound and CGE as a lower bound of impact).

Another methodology is to use historical data to statistically infer economic impacts. One insurance claim study is a prime example, using regression analysis to link geomagnetic indices with observed equipment failure claims (Schrijver, 2015). Similarly, researchers could examine macroeconomic time series (industrial output, electricity sales, etc.) around known solar storms to estimate deviations attributable to specific events (Forbes and St. Cyr, 2019). So far, empirical economic analyses of space weather are sparse due to limited events and confounding factors. However, this approach is common in climate and weather impact economics, and can be applied when enough data accrue, for instance, isolating the effect of space weather on monthly electricity reliability metrics. Statistical approaches can capture subtler, chronic costs of space weather (like the few-percent annual loss noted by Schrijver) that models might miss (Forbes and St. Cyr, 2004; Forbes and Cyr, 2008).

To tie together the above sections, it is useful to consider how a major space weather event's impacts would be assessed in practice, using a combination of historical analogues and modelling. Imagine a major geomagnetic disturbance that knocks out power to a large portion of a continent for multiple days. Analysts would proceed on multiple fronts. Firstly, in estimating direct infrastructure damages such as counting how many transformers might fail and the cost to replace them (each large EHV transformer can cost $1–10 million). Secondly, calculating immediate economic losses from outages, for example, using VoLL for the hours of electricity not delivered to millions of customers, yielding a base dollar loss. Thirdly, modelling

the indirect macroeconomic effects, such as via IO simulation, to capture lost production in manufacturing, reduced consumer spending, etc., which might double the loss figure over the outage period. Fourthly, incorporating restoration duration, as the outage is likely to last multiple days or weeks. Finally, results can be validated against analogous events, such as by comparing magnitudes of computed losses to known events like the 2003 blackout (US$4–10 billion) or the 1989 Quebec outage (a few hundred million) to ensure consistency in scaling. Through such a method, an overall estimate emerges of the potential economic impact of such an event.

## 3. Method

We develop a coupled physics-engineering-economic spatial modelling framework, capable of answering the research questions specified in the introduction, to assess the economic impact of power outages from geomagnetic disturbances. This involves estimating economic impacts using four complementary methods designed to provide a consistent comparison between demand-side and supply-side shock formulations. First, we apply a demand-side Leontief input-output model in which disruption is represented as a population-weighted reduction in household final demand. Second, we apply a demand-side Leontief model in which disruption is weighted using survey-based household Value of Lost Load (VoLL) estimates derived from Transpower substation data. Third, we apply a supply-side Ghosh model with percentage shocks representing proportional reductions in sectoral productive capacity. Fourth, we apply a supply-side Ghosh model using survey-based VoLL-derived direct losses. This four-method structure allows direct comparison across scenarios while preserving parallel demand-side and supply-side interpretations under both percentage and survey-based shock assumptions.

### 3.1. The Aotearoa New Zealand context

The geographical and geophysical characteristics of Aotearoa New Zealand highlight the need for assessing GIC vulnerability. The country spans geomagnetically active mid- to high-latitude bands, with the North Island located between approximately 34° and 42°S, and the South Island stretching from 41° to 47°S (Mukhtar *et al.*, 2020). These latitudes are within the range where GICs are known to pose increased risk during geomagnetic storms due to substantive magnetic field perturbations (Divett et al., 2020).

Compounding this is the variable nature of ground conductivity, which plays a significant role in influencing the intensity and distribution of GICs during space weather events (Divett *et al.*, 2017). In particular, areas with low conductivity (i.e., high resistivity) can enhance induced electric fields, causing GICs to flow into EHV transformers and other grid components causing increased stress (Pratscher *et al.*, 2024). Currently, magnetotelluric (MT) measurements have been made at 62 sites in the southern part of South Island to further scientific understanding of this contributing factor (Ingham *et al.*, 2023).

The present New Zealand GIC model considers four key network characteristics which include, (i) the location of substations and their earthing status via the associated transformers primary windings, (ii) transmission line connections between substations, (iii) transformer quantity and types at substations including their primary winding earthed/unearthed status, and (iv) electrical resistance values for substations, transformers, and transmission lines (Mac Manus *et al.*, 2022). In terms of assets, the national grid infrastructure includes high voltage transmission lines operating at 220 kV and 110 kV, supported by a relatively small number of EHV transformer units, as illustrated in Figure 1.

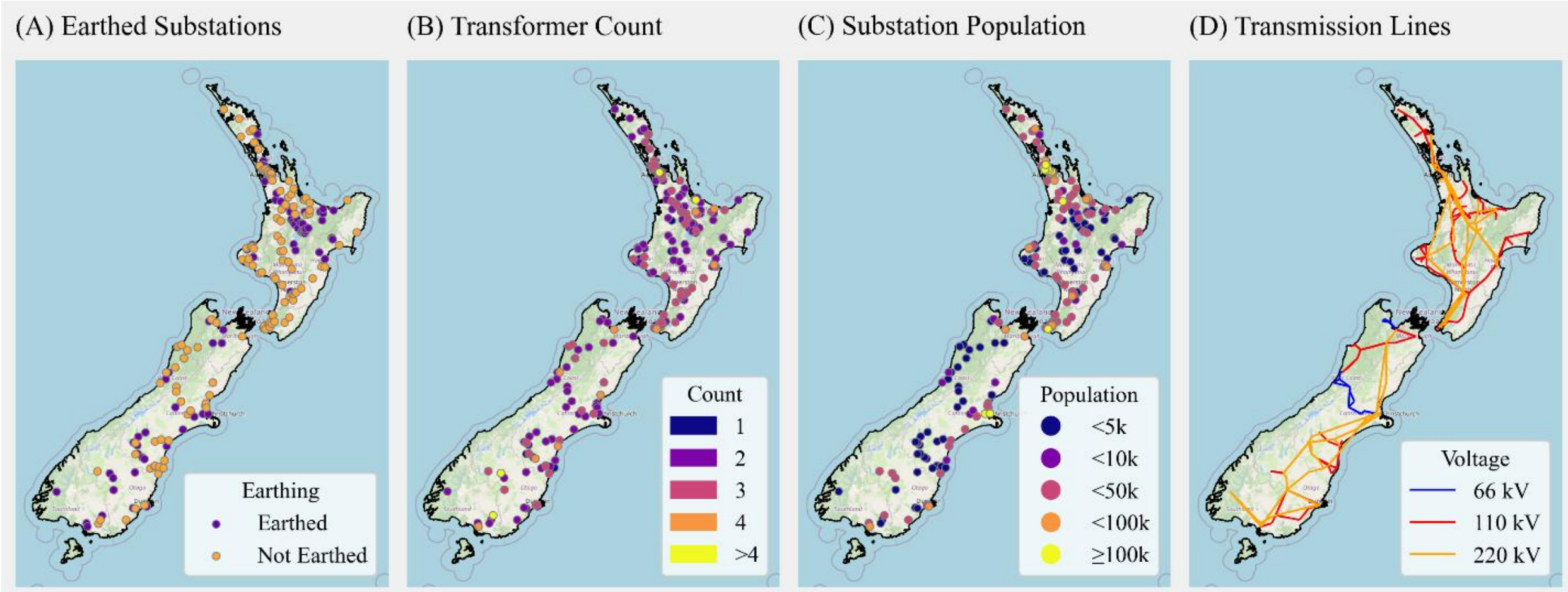

Figure 1 Context of Aotearoa New Zealand's electricity transmission infrastructure

Subplot (A) maps the earthing status of substations[1], showing a mix of earthed and non-earthed facilities distributed throughout both the North and South Islands. Subplot (B) illustrates the number of transformers per substation, revealing that most substations house one or two transformers, while a smaller number contain more than four (mainly in high-demand areas). Subplot (C) displays the population served by each substation, indicating that the majority serve fewer than 10,000 people, though several substations in cities such as Auckland, Wellington and Christchurch serve populations exceeding 100,000. Subplot (D) shows the high-voltage transmission network, distinguishing by voltage level, specifically for 66 kV, 110 kV, and 220 kV lines. The 220 kV lines constitute the primary transmission backbone, spanning both islands, while lower-voltage lines form denser regional networks. Together, these plots provide an integrated contextual understanding of the present EHV electricity transmission network.

### 3.2. Scenarios

Seven initial scenarios are stated in Table 1, which reflect a variety of different potential future outcomes. The first two scenarios (1 & 2) are basic in their approach and represent the worst-case estimates to provide an upper bound to the analysis. No long duration outages are expected in any scenario (e.g., months) as the network owner would remove exposed transformers from service, prior to being catastrophically impacted. Scenario 1 is a total loss of power for both islands over six days due to no mitigation investments having been made. Power is fully lost for three days (seen as the longest duration of the event) and then returned via a restoration sequence which begins at the hydro generation sites inland on the North and South Islands, moving outwards one substation at a time radially along network lines (based on existing operational plans).

By contrast, Scenario 2 represents a loss of power in the South Island of New Zealand over six days, following the same restoration approach as the prior scenario. Due to the loss of the High Voltage Direct Current (HVDC) interconnector between the North and South Islands (transferring power from southern generation sites to the more populated north), the North Island undergoes 20% load shedding over six days, until power is fully restored and the HVDC link functioning again.

[1]All substations have an earth grid and are earthed for safety. `Earthed status` refers to the effective earthing of the primary power system at the specific substation via the transformers installed at a site. Should a site have transformers with a Delta-Star (D-y) configuration of primary and secondary windings, it is considered non-effectively earthed as it relies on the connecting lines to provide a "Remote" earth reference. Where a site has transformers with Star-Star (Y-y) or Star-Delta (Y-d) configuration of primary and secondary windings, it is considered effectively earthed as an earth reference is provided at the site.

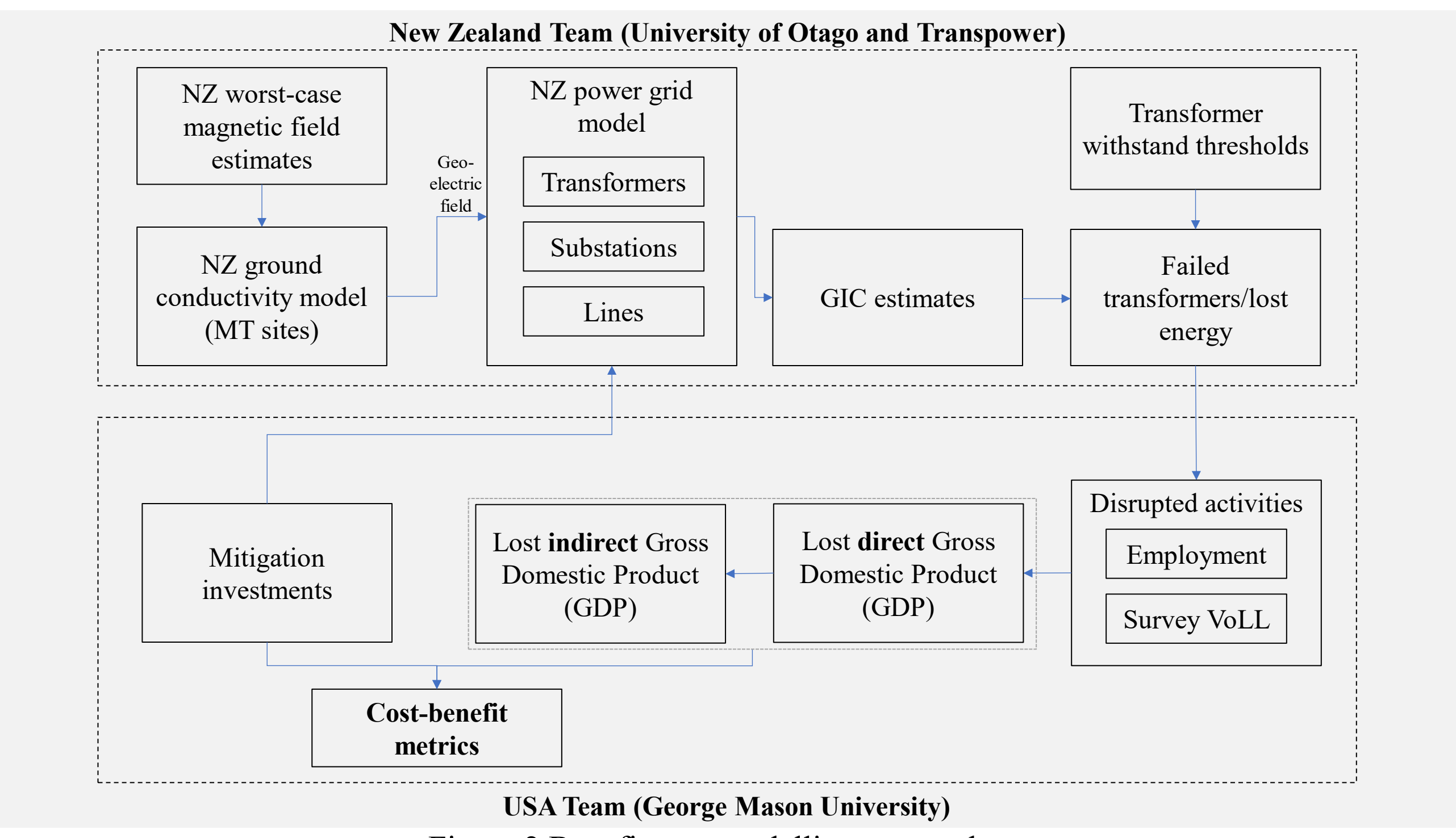

Figure 2 Benefit-cost modelling approach

By contrast, the next five scenarios utilize estimated GIC values, most recently developed from the Solar Tsunami Endeavour Programme utilizing the 1-minute peak and 60-minute mean GIC (A) exposure, and transformer withstand thresholds defined by the team. For scenarios 4-7, mitigation is assumed to be implemented prior to storm commencement. Initially, Scenario 3 considers no mitigating actions, where substations exposed to >500 A are removed from service, based on two transformers at each site being exposed evenly to 250 A GIC in their neutral (~83A/ph) each, and either being purposefully removed from service to avoid damage, or protective relay equipment bringing each asset offline at an expected 130 ℃. A similar restoration process is adopted to Scenarios 1 & 2, whereby there is a full power outage from days one to three while the storm passes, with restoration then taking up to another three days. Power is returned based on the restoration sequence developed which accounts for inland hydro generation sites, with power moving outwards along the radial network lines from these sites. In such a situation, it is expected that during a winter peak period North Island would experience 20% load shedding due to a lack of South Island generation imports, returning to full power by the end of the restoration period as the HVDC interconnect is returned to service.

| Scenario | Complexity | Spatial impact | Temporal duration | Restoration curve | Investment |
|---|---|---|---|---|---|
| **1** | Basic | North and South Island | 6 days | Restoration from day 4 via hydro generation sites. | None |
| **2** | Basic | South Island treated the same as Scenario 1. North island has 20% load shedding due to loss of HVDC interconnector. | 6 days | Restoration from day 4 for South Island via hydro generation sites. 6 days of load shedding in North Island. | None |
| **3** | Advanced, using original baseline current estimate | Based on breached transformer thresholds from NZ GIC Model, where substations >500 A GIC fail. North Island sees 20% load shedding. | | Restoration from day 4 for South Island over 3 days via hydro generation sites. 6 days of load shedding in North Island. | None |
| **4** | Advanced, using mitigation switching sequence | Based on breached transformer thresholds from NZ GIC Model, where substations >500 A GIC fail. North Island sees 20% load shedding. | | Restoration from day 4 for South Island over 3 days via hydro generation sites. 6 days of load shedding in North Island. | SW R&D to develop switching sequence ($250k) |
| **5** | Advanced, using mitigation switching | Based on breached transformer thresholds from NZ GIC Model, where substations >500 A GIC fail. North Island sees 20% load shedding. | | Restoration from day 4 for South Island over 3 days via hydro generation sites. 6 days of load shedding in North Island. | SW R&D to develop switching + islanding sequence ($500k) |

| | sequence plus islanding | | | |
|---|---|---|---|---|
| **6** | Advanced, using mitigation switching sequence and limited blocker deployment | Based on breached transformer thresholds from NZ GIC Model, where substations >500 A GIC fail. North Island sees 20% load shedding. At each site, 1 site transformer blocked, 1 switched off. | Restoration from day 3 for South Island over 2 days via hydro generation sites. 4 days of load shedding in North Island. | SW R&D to develop switching sequence and islanding ($750k) + Blockers (12 blocked @ NZ$2m each = $24m) |
| **7** | Advanced, using mitigation switching sequence plus islanding, and enhanced blocker deployment | Based on breached transformer thresholds from NZ GIC Model, where substations >500 A GIC fail. North Island sees 20% load shedding. At each site, all transformers blocked. | Restoration from day 3 for South Island over 1 day via hydro generation sites. 3 days of load shedding in North Island. | SW R&D to develop switching sequence and islanding ($750k) + Blockers (34 blocked @ NZ$2m each = $68m) |

Table 1 Scenarios defined in collaboration with the electricity transmission operator Transpower

Scenario 4 applies the same spatial exposure framework, failure threshold, restoration logic, and North Island load-shedding assumption as Scenario 3, but uses the post-switching GIC estimates to represent the effect of the mitigation switching sequence. The investment cost for this scenario is NZ$250k representing seven engineers working over four weeks to develop the switching sequence.

Next, Scenario 5 consists of both the developed switching sequence, as well as additional islanding of the strategically important New Zealand Aluminium Smelters (NZAS) plant at Tiwai Point (TWI) (and others), with similar asset vulnerability, restoration and load shedding (NZAS supports a multi-billion-dollar export market). Islanding is the process of segmenting a portion of the electrical grid that can operate independently from the wider grid. In this scenario, we assume the TWI substation has agreed to block two transformers, allowing Transpower to isolate the substation to Manapouri (MAN) by reconfiguring the circuits between MAN, North Makarewa (NMA), and Invercargill (INV) into a point-to-point island. This ensures all MAN generation directly supplies the TWI load, protecting the smelter plant. The investment cost for this scenario is NZ$500k representing seven engineers working over four weeks to develop the switching sequence, plus four weeks to develop the islanding strategy.

Two additional scenarios are explored which include supplementary investments, for example, in GIC blocking devices (totalling NZ$24.75 & 68.75 million) (as visualized in Figure 3). In Scenario 6, a mitigation switching sequence is implemented to reduce GIC across the network. An islanding scheme was created for TWI and MAN, with blockers installed at T1 and T2, resulting in zero GIC through both substations. NMA also sees T1 blocked and T2 disconnected, eliminating GIC flow through the site. INV sees T3 and T5 blocked and T1 disconnected, achieving the same effect. In South Dunedin (SDN) and Halfway Bush (HWB), T2 at SDN and T3 and T6 at HWB are disconnected to stop GIC penetration. Gore (GOR) has T11 blocked and T12 disconnected, while Cromwell (CML) has T8 blocked. Timaru (TIM) undergoes disconnection at T5 and blocking at T8, effectively stopping GIC. At Islington (ISL), transformers T1, T6, and T8 are blocked and T2, T3, and T7 are disconnected, isolating the substation from GIC. Similarly, Bromley (BRY) has T5 and T7 blocked. Further measures include disconnecting transmission lines south of Kikiwa (KIK), isolating KIK, Stoke (STK), Argyle (ARG), and Cobb Valley (COB) as an island. Additional disconnections west of Kikiwa and Hororata (HOR) separate Murchison (MCH), Inangahua (IGH), Robertson Street (ROB), Reefton (RFN), Atarau (ATU), Dobson (DOB), Greymouth (GYM), Kumara (KUM), Hokitika (HKK), Otira (OTI), Arthurs Pass (APS), Castle Hill (CLH), and Coleridge (COL), forming another electrically isolated region to mitigate GIC impacts. The investment costs involve NZ$750k labour for the switching sequence and to install blockers at 12 substations, with the blocker equipment costing NZ$24 million (NZ$2 million each).

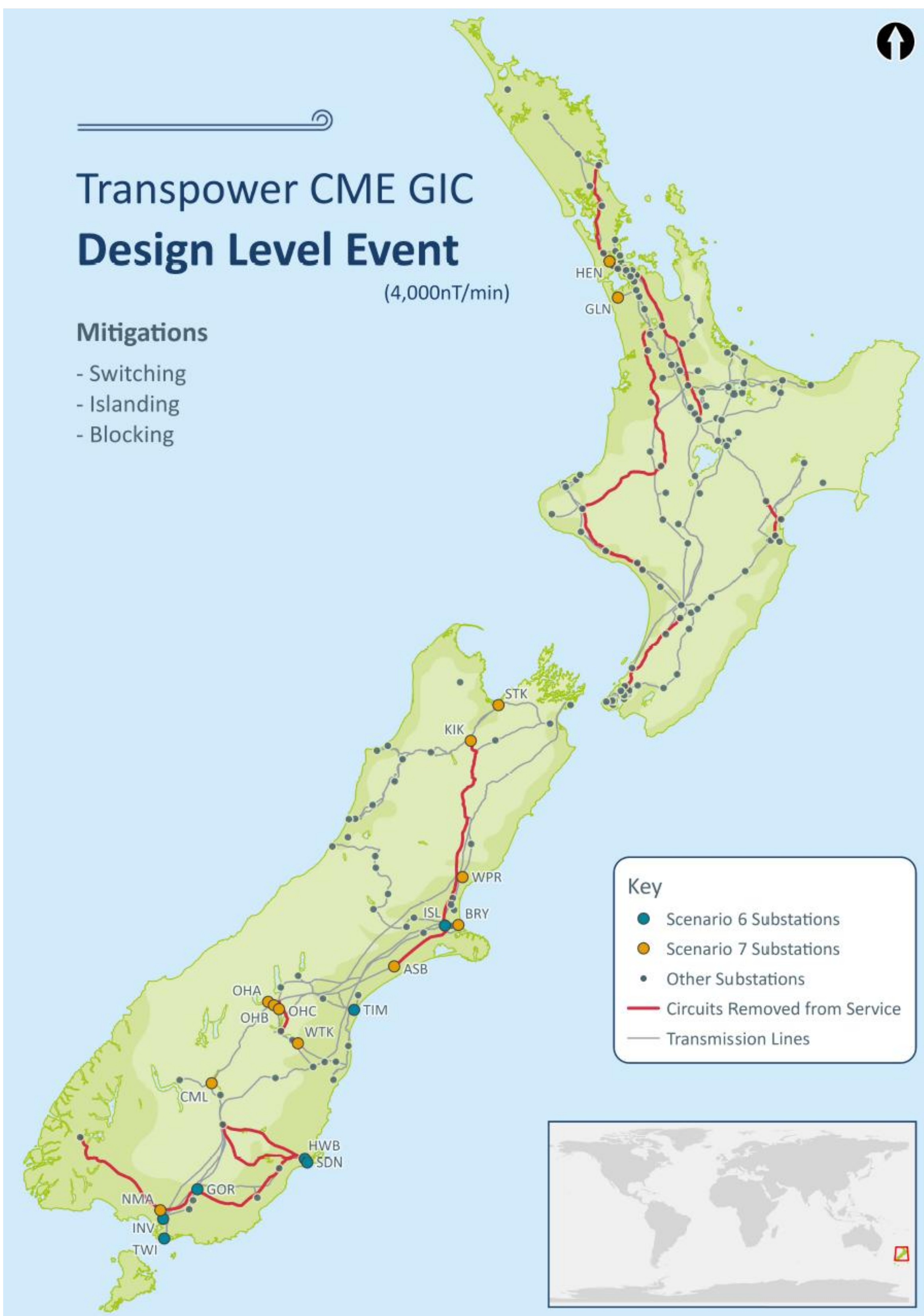

*Figure 3 Transpower grid with blocked substation locations for scenarios 6 and 7*

Finally, in Scenario 7 additional blockers are added beyond the installations in the previous scenario. For example, CML sees T5 also blocked, followed by transformers T1, T2, T3, and T5 at Henderson (HEN). Glenbrook (GLN) is similarly protected by blocking T4, T5, and T6. At Ashburton (ASB), a broader set of transformers (T1, T3, T8, T9, and T10) are blocked, resulting in complete mitigation of GIC. Waipara (WPR) sees T12 and T13 blocked, while KIK is secured through blocking of T1 and T2. Stoke (STK) undergoes extensive transformer blocking at T3, T4, T6, T7, and T10. Mitigation is also extended to key hydro generation sites, including Ohau A (OHA) with T4 through T7 blocked, Ohau B (OHB) with T8 through T11, and Ohau C (OHC) with T12 through T15. Collectively, these measures ensure zero GIC flow through all targeted substations and generation assets, enhancing system resilience against geomagnetic disturbances. The investment costs involve NZ$750k labour for the switching sequence and to install blockers at the previous 12 substations, plus an extra 22 sites (34 in total), with the blocker equipment costing NZ$68 million (NZ$2 million each).

### 3.3. Method - Restoration

This section makes specific reference to the spatio-temporal progression of power outage restoration across the presented scenarios following a hypothetical large-scale disruption in Aotearoa New Zealand. For a graphical understanding of the restoration process, see Figures S1 and S2 in the supplementary information. In Scenario 1, the entire country encompassing both the North Island and the South Island, experiences widespread outages on Day 1, with restoration resuming gradually from day three from hydro generation sites inland (see Figure S2). Major metropolitan areas such as Auckland, Wellington, and Christchurch regain power by Day 5, and near-complete restoration achieved by Day 6.

Scenario 2 displays a contrasting geographic pattern, with the North Island remaining unaffected while the South Island undergoes a prolonged blackout. Recovery in this case is similarly incremental, with the northern half of the South Island regaining power by Day 4 and substantial restoration observed across the island by Day 6.

Scenarios 3 through 7 depict more localized outages using the more nuanced GIC modelling from the New Zealand GIC model, leading to more targeted disruption (see Figure S2). In Scenario 3, initial power loss is concentrated in the central South Island, with power returning incrementally across adjacent regions by Day 6. Scenario 4 demonstrates a similar spatial pattern, but with slightly delayed restoration along the eastern coastline, such as around Napier, and with most regions showing recovery by Day 5. Scenario 5 indicates more isolated disruption in the lower South Island and southern North Island, thanks to the combination of switching and islanding. With a smaller disruption zone, restoration here is more rapid, with most regions showing recovery by Day 5.

Finally, in the two blocker scenarios, Scenario 6 reveals a more dispersed outage pattern, with early impacts observed in scattered pockets across both islands, particularly in the more rural parts of the inland Otago region. Due to the smaller number of disrupted areas, power is completely restored by the end of Day 4. Scenario 7 presents the most limited and rapidly resolved outage, with minor disruptions mostly confined to the southern South Island (except for Napier) due to enhanced blocker installation, allowing close to full restoration by the end of Day 3. These restoration sequences recognize that reduced GIC vulnerability enables more rapid restoration, with power being restored in as little as 24-48 hrs after the storm has passed. This is because all available resources can be targeted at a smaller number of vulnerable assets.

In the model code, scenario preprocessing generates daily outage indicators for days 1–7, with day 7 representing full restoration. The economic loss calculations aggregate disruption over days 1–6, consistent with the six-day impact horizon used in the reported results.

### 3.4. Direct economic impacts

To estimate the direct impact of a severe space weather event, we first translate outage exposure into disrupted electricity consumption at the local and sectoral level. As detailed macroeconomic output data are not available at the same spatial resolution as the modelled outage areas, we use employment and population data from Stats New Zealand as a proxy for the spatial distribution of economic activity (Stats NZ, 2024). This approach allows national electricity consumption to be downscaled to local transmission node service areas and then re-aggregated into industrial sector and household disruption estimates. This also ensures that all subsequent economic calculations are based on the same spatial representation of outage exposure.

For each industrial sector $j$, we calculate electricity intensity per employee ($Electricity\ Intensity\ Per\ Employee_j$) (in MWh) as the ratio of total sectoral electricity use to total sectoral employment, as shown in equation (1). This provides a sector-specific estimate of average electricity demand per worker and is used to translate disrupted employment into disrupted electricity use.

$$Electricity\ Intensity\ Per\ Employee_j = \frac{Total\ Electricity\ Use_j}{Employment_j} \tag{1}$$

This is based on dividing the total electricity used by each industrial sector ($Total\ Electricity\ Use_j$) (also in MWh), by the associated employment ($Employment_j$). See supplementary information Figure S3 to view the real data graphically, where subplot (A) illustrates electricity use across major economic sectors in New Zealand using statistics from the New Zealand Energy Quarterly dataset (Ministry of Business, Innovation & Employment, 2024). The Commercial sector dominates electricity use with over 9,200 GWh annually, followed by Basic Metals (>5,600 GWh) and Food Processing (3,100 GWh). In contrast, sectors such as Transport (277 GWh) and Chemicals (427 GWh) consume significantly less electricity, reflecting both lower energy intensity and reliance on alternative energy sources (in the case of Transport). Figure S3 (B) also presents the employment by sector where the Commercial segment dominates with 1.81 million employees, dropping substantially for other more minor industrial groupings. Moreover, Figure S3 (C) visualises the electricity consumption per employee, which can then be used to estimate local disrupted electricity use ($Local\ Electricity\ Use_{l,j,t}$) for each sector $j$ in each local area $l$ and time period $t$ by multiplying electricity intensity per employee ($Electricity\ Intensity\ Per\ Employee_j$) by disrupted local employment ($Disrupted\ Local\ Employment_{l,j,t}$), as shown in equation (2).

$$\begin{aligned} &Local\ Electricity\ Use_{l,j,t} \\ &\quad = Electricity\ Intensity\ Per\ Employee_j \\ &\quad \times Disrupted\ Local\ Employment_{l,j,t} \end{aligned} \tag{2}$$

The level of disrupted employment ($Disrupted\ Local\ Employment_{l,j,t}$) in each local sector is obtained using the restoration scenarios already outlined. Finally, we can sum this across all disrupted local areas $l$ and time periods $t$ to obtain the lost load ($Lost\ Load_j$) nationally for each $j$ industrial sector, as in equation (3)

$$Lost\ Load_j = \sum_l \sum_t Local\ Electricity\ Use_{l,j,t} \tag{3}$$

We then translate lost electricity into estimated monetary loss using Value of the Lost Load ($VoLL_j$) for each $j$ sector, as shown in equation (4). In the industrial sector VoLL formulation, this produces a direct economic loss estimate based on the opportunity cost of unserved electricity. These values are conservative because they represent the value of interrupted electricity service rather than the full economy-wide consequences of disrupted production and supply chains.

$$Direct\ Economic\ Loss_j = Lost\ Load_j \times VoLL_j \quad (4)$$

This estimates the direct economic loss of this disruption by $j$ sector ($Direct\ Economic\ Loss_j$) that can subsequently be translated into total impacts using the alternative economic estimation approaches. Indeed, the direct estimates derived in this section are used in two different ways in the four-method framework. For the Demand-Side Leontief (Population-Weighted Shock) method, the physical outage scenarios are translated into a reduction in household final demand based on the share of population disrupted. For the Demand-Side Leontief (Survey-Based VoLL Shock) method, disrupted population is weighted using household VoLL estimates derived from Transpower survey data before being aggregated into a scenario-level household final demand shock. For the Supply-Side Ghosh (Employment-Weighted Shock) method, the same physical disruption data are converted into proportional sectoral value added shock factors based on disrupted employment shares. Finally, for the Supply-Side Ghosh (Survey-Based VoLL Shock) method, VoLL-derived direct sectoral losses using Transpower survey data are used as the exogenous value added shocks that enter the macroeconomic model.

The resulting lost load and direct loss estimates provide the common physical outage baseline for the four economic impact methods. The two Ghosh methods translate these outages into sectoral supply shocks, either using disrupted employment shares or VoLL-derived direct losses. The two Leontief methods translate the same outages into reductions in household final demand, either using disrupted population shares or normalized household VoLL-weighted disruption factors. The methods use a common outage baseline while differing only in how those outages are translated into economic losses, aligning with established approaches in the literature (Hashemi *et al.*, 2018; Hashemi, 2021; Thomas and Fung, 2022; Jin, Lee and Kim, 2023; Sprintson and Oughton, 2025).

Current VoLL estimates are also presented by sector in the supplementary information Figure S3 (D) for context. VoLL represents the economic cost of unserved energy and varies widely across sectors. The Transport sector exhibits the highest VoLL at $55,941/MWh, because it consumes only a very small amount of electricity currently (277 GWh).

### 3.5. Total and indirect economic impacts

To calculate the total and indirect macroeconomic impact, national account data are obtained from the New Zealand statistics authority to support development of a consistent sector-by-sector representation of inter-industry linkages (Stats NZ, 2021). IO models are particularly useful for assessing infrastructure disruption because they capture how losses in one part of the economy propagate through upstream suppliers, downstream customers, and final demand. In this study, we use two IO structures, each paired with two shock specifications, yielding four economic methods in total, including the Demand-Side Leontief (Population Shock), Demand-Side Leontief (Survey-Based VoLL), Supply-Side Ghosh (Employment Shock), and Supply-Side Ghosh (Survey-Based VoLL).

For the two demand-side methods, indirect economic impacts are estimated using a standard Leontief input-output model:

$$\mathbf{x} = (\mathbf{I} - \mathbf{A})^{-1}\mathbf{f} \tag{5}$$

where $\mathbf{x}$ is the gross output vector, $\mathbf{A}$ is the Leontief technical coefficients matrix, $\mathbf{I}$ is the identity matrix, $\mathbf{f}$ is the final demand vector, and $(\mathbf{I} - \mathbf{A})^{-1}$ is the Leontief inverse. Vectors are expressed as column vectors. For each outage scenario $s$, household final demand is reduced using a scenario disruption factor:

$$\mathbf{f}^s = \mathbf{f}^0 - \theta^s\,\mathbf{h} \tag{6}$$

where $\mathbf{f}^0$ is the baseline final demand, $\mathbf{f}^s$ is the shocked final demand, $\mathbf{h}$ is the household final demand vector, and $\theta^s$ is a dimensionless proportional reduction in household demand. In the population shock Leontief method, $\theta^s$ reflects the share of the disrupted population. Under the survey-based household VoLL Leontief method, $\theta^s$ reflects the VoLL-weighted household disruption factor derived from household interruption cost estimates from the Transpower survey, allowing higher cost outage areas to contribute more strongly to the aggregate household demand shock. Demand-side gross output losses are then calculated as:

$$\mathbf{L}^s = (\mathbf{I} - \mathbf{A})^{-1}(\mathbf{f}^0 - \mathbf{f}^s) \tag{7}$$

where $\mathbf{L}^s$ is the vector of sectoral gross output losses under scenario $s$. When losses are reported in GDP terms, gross output losses are converted to value-added losses using sector-specific value-added to output ratios.

For the two supply-side methods, indirect impacts are estimated using a Ghosh allocation model. The Ghosh coefficient matrix ($\mathbf{B}$) is produced by normalizing inter-industry sales by the total output of the supplying sector:

$$b_{ij} = \frac{z_{ij}}{x_i} \tag{8}$$

where $z_{ij}$ is the sale from sector $i$ to purchasing sector $j$, and $x_i$ is the total output of supplying sector $i$. For the Ghosh allocation model, value-added and output vectors are expressed as row vectors, consistent with the row-normalized allocation matrix $\mathbf{B}$.

$$\mathbf{G} = (\mathbf{I} - \mathbf{B})^{-1} \tag{9}$$

For the Supply-Side Ghosh (% Shock) method, shocked value added is calculated as:

$$VA_j^s = VA_j\,(1 - \phi_j^s) \tag{10}$$

where $VA_j^s$ is the shocked value added for each sector $j$ under scenario $s$, and $VA_j$ is the baseline value added for sector $j$, and $\phi_j^s$ is the proportional disruption factor derived from disrupted employment shares.

For the Supply-Side Ghosh (Survey-Based VoLL) method, shocked value added is calculated as:

$$VA_j^s = \max(0, VA_j - DL_j^s) \quad (11)$$

where $DL_j^s$ is the VoLL-derived direct economic loss for sector $j$ under scenario $s$. The shocked value-added row vector, $\mathbf{va}^s = [VA_1^s, VA_2^s, \dots, VA_n^s]$, is then propagated through the Ghosh inverse to estimate post-shock output:

$$\mathbf{x}^s = \mathbf{va}^s\, \mathbf{G} \quad (12)$$

where $\mathbf{x}^s$ is the post-shock gross output row vector under scenario $s$. Total gross output losses are calculated as the difference between baseline and post-shock output. These gross output losses are then converted to equivalent GDP losses using sector value-added to output ratios. Indirect losses are then calculated as total GDP losses minus the corresponding direct GDP losses, ensuring consistent units when separating direct and indirect components.

Finally, economic impact estimates derived from March 2020 New Zealand input-output tables are converted to 2026 New Zealand dollars using an implicit GDP price deflator. The scalar is calculated as the ratio of the expenditure-based GDP implicit deflator in the year ended March 2026 to that in the year ended March 2020, using nominal and real GDP series from RBNZ/Stats NZ. This gives a conversion factor of 1.255. Thus, all 2020 NZ dollar results are multiplied by 1.255 to express them in 2026 NZ dollars.

## 4. Results

This section reports the results of the seven disruption and mitigation scenarios using the four different economic impact estimation methods.

### 4.1. Direct and indirect GDP impacts

Figure 4 presents the estimated direct and indirect GDP losses under the seven different scenarios, reported in billions of New Zealand dollars, whereas Figure 5 reports impacts as a percentage of annual GDP. In Scenario 1 we see the highest total GDP loss across all methods, reflecting the national nature of the potential disruption from space weather. Using the Supply-Side (Employment Shock), total GDP losses reach NZ$3.58 billion (0.98% of GDP) as an upper bound across the full set of scenarios. The Supply-Side Ghosh (Survey-Based VoLL Shock) method produces a lower but still substantial estimate of NZ$2.21 billion (0.6% of GDP). By contrast, the two demand-side Leontief approaches reach similar quantities, with total GDP losses of NZ$1.75 billion (0.48% of GDP) under the population-weighted shock and NZ$1.76 billion (0.48% of GDP) under the household survey-based VoLL shock. This indicates that although methodological choice affects the absolute level of estimated loss, all four approaches identify Scenario 1 as the most economically damaging case. By contrast, Scenario 2 sees lower impacts of NZ$0.75, NZ$0.75, NZ$1.49, and NZ$0.94 billion, equivalent to 0.20%, 0.20%, 0.41%, and 0.26% of annual GDP, for the Demand-Side (Population Shock), Demand-Side (Survey-Based Shock), Supply-Side (Employment Shock), and Supply-Side (Survey-Based Shock), respectively.

Scenario 3 provides the main GIC baseline where no mitigation has been applied. Here, total GDP losses are broadly similar to Scenario 2, with demand-side estimates of NZ$0.74–0.76 billion (0.2-0.21% of GDP) and supply-side estimates of NZ$0.93–1.48 billion (0.25-0.41% of GDP). The Supply-Side Ghosh (Employment Shock) method again produces the largest estimate, while the Supply-Side Ghosh (Survey-Based VoLL Shock) method provides an intermediate estimate. Compared to Scenario 1 and 2, this scenario demonstrates that even when outage areas are more spatially targeted, economic impacts remain substantial due to indirect supply chain propagation.

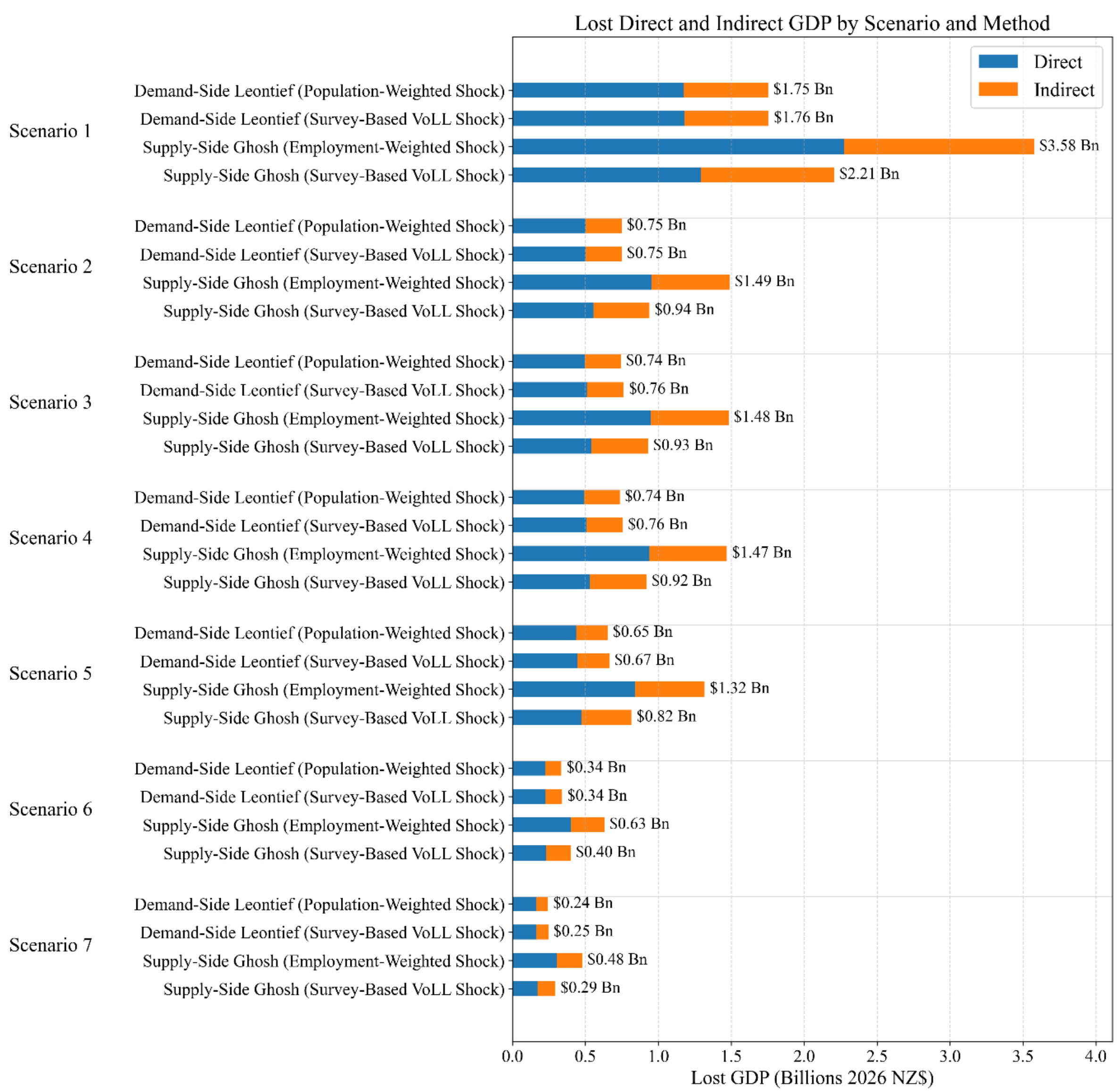

Figure 4 Comparison of modelled GDP impact estimates

Next, the mitigation switching sequence is implemented in Scenario 4, producing a modest reduction in GDP losses relative to Scenario 3. Total losses are estimated at NZ$0.74 billion (0.2% of GDP) under the population-weighted Leontief method, NZ$0.76 billion (0.21% of GDP) under the survey-based household VoLL Leontief method, NZ$1.47 billion (0.4% of GDP) under the employment-weighted Ghosh method, and NZ$0.92 billion (0.25% of GDP) under the survey-based VoLL Ghosh method. Thus, while the switching sequence reduces the number or severity of disrupted assets relative to the baseline GIC case, the aggregate economic impact remains close to Scenario 3 under the central modelling assumptions.

By contrast, when introducing islanding on top of the switching sequence in Scenario 5, substantial losses are avoided. Total GDP losses fall to NZ$0.65–0.67 billion (0.18% of GDP) under the two demand-side

Leontief methods. Under the supply-side approaches, losses are reduced to NZ$1.32 billion (0.36% of GDP) using the employment-weighted Ghosh method and NZ$0.82 billion (0.22% of GDP) using the survey-based VoLL Ghosh method.

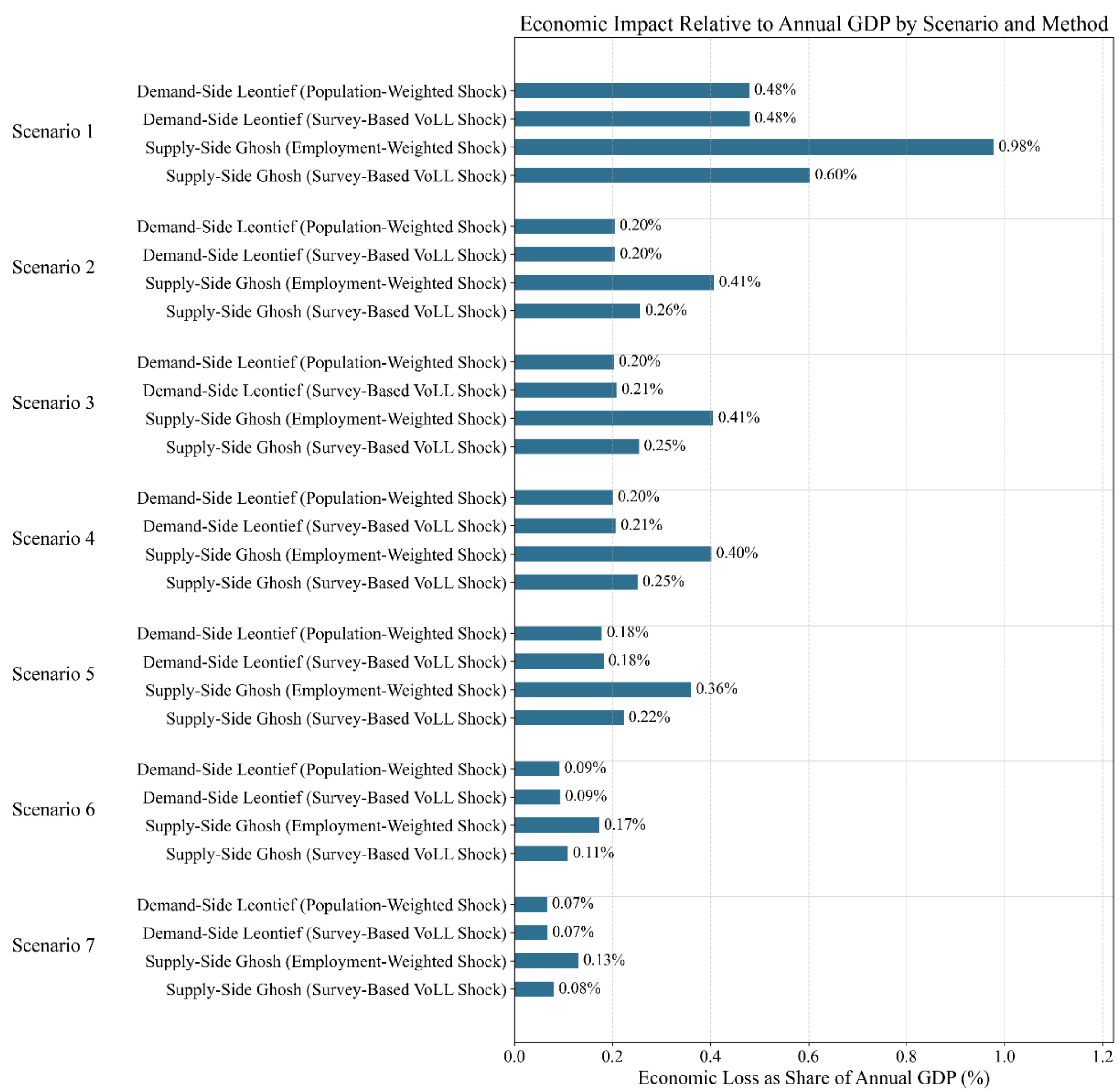

Figure 5 Comparison of modelled impacts as a percentage of annual GDP

Finally, the blocker deployment variants (Scenarios 6 and 7), produce the lowest economic losses. In Scenario 6, which includes limited GIC blocker deployment, total GDP losses fall to NZ$0.34 billion (0.09% of GDP) under both demand-side Leontief methods, NZ$0.63 billion (0.17% of GDP) under the employment-weighted Ghosh method, and NZ$0.40 billion (0.11% of GDP) under the survey-based VoLL Ghosh method. Scenario 7, which represents enhanced blocker deployment, reduces losses further, with total GDP impacts of NZ$0.24–0.25 billion under the demand-side approaches (0.07% of GDP), NZ$0.48 billion (0.13% of GDP) under the employment-weighted Ghosh approach, and NZ$0.29 billion (0.08% of GDP) under the survey-based VoLL Ghosh approach.

## 4.2. Benefit-cost ratios of mitigation investment

Figure 6 presents the estimated benefit-cost ratios for Scenarios 4–7, using Scenario 3 as the no-mitigation GIC-model baseline. Scenarios 1–3 are treated as having no mitigation investment and therefore do not receive benefit-cost ratios. For the mitigation scenarios, benefits are calculated as the avoided GDP loss relative to Scenario 3, while costs reflect the estimated expenditure required to implement the relevant operational or hardware-based mitigation strategy.

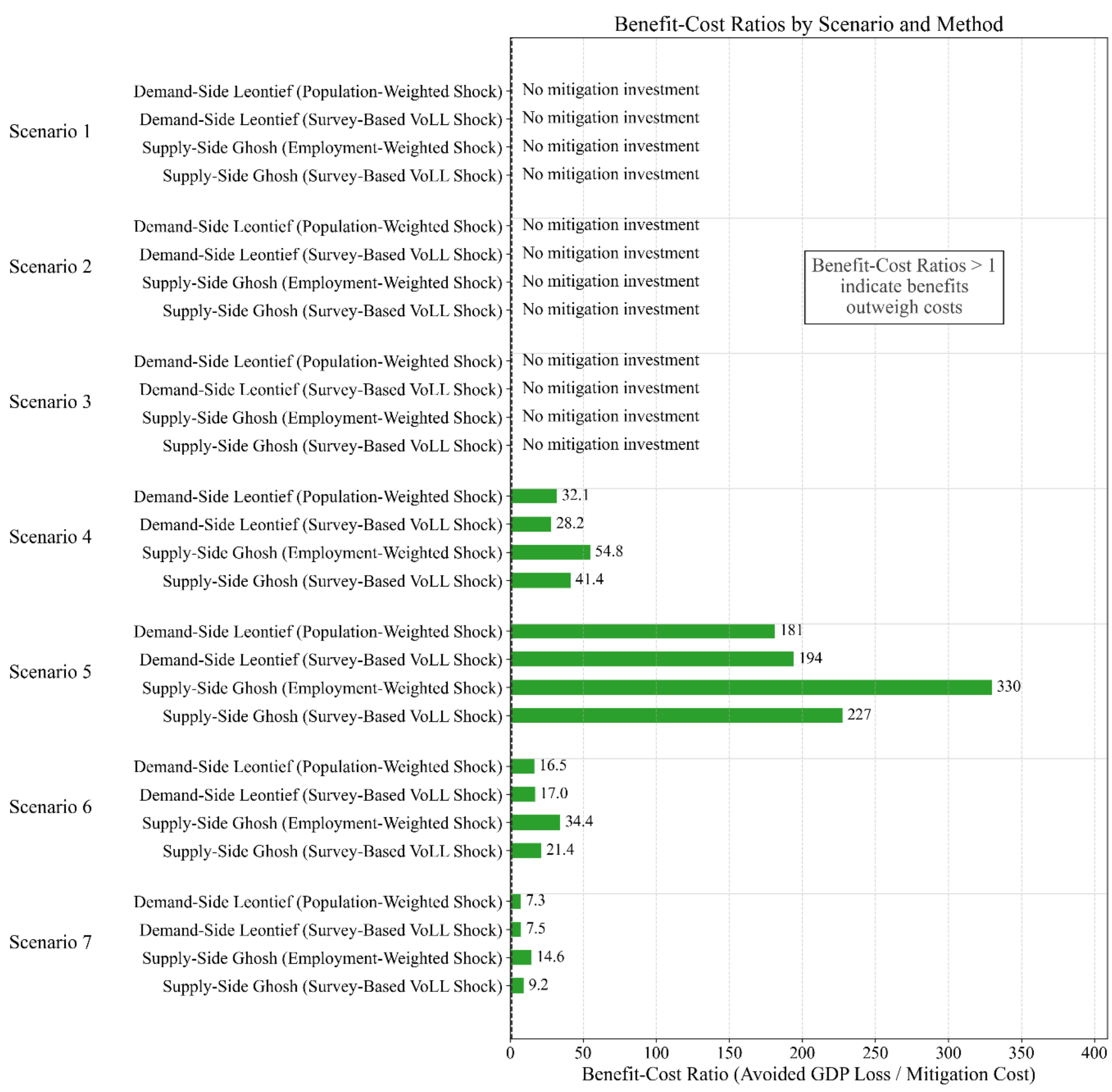

Figure 6 Comparison of Benefit-Cost Ratio (BCR) estimates

Scenario 4, which implements the mitigation switching sequence at an estimated cost of NZ$250,000, produces positive benefit-cost ratios across all four economic methods. The benefit-cost ratio ranges from 28.2 under the demand-side Leontief survey-based VoLL method to 54.8 under the employment-weighted Ghosh method. The population-weighted Leontief and survey-based Ghosh methods produce ratios of 32.1

and 41.4, respectively. These values demonstrate that even modest avoided losses can justify investments in lower cost operational planning investments.

By contrast, in Scenario 5 the highest benefit-cost ratios are achieved across all mitigation options. This scenario combines the switching sequence with islanding, at an estimated investment cost of NZ$500,000. Benefit-cost ratios range from 181 under the population-weighted Leontief method to 330 under the employment-weighted Ghosh method. The survey-based household VoLL Leontief and survey-based VoLL Ghosh methods produce benefit-cost ratios of 194 and 227, respectively.

Finally, in Scenarios 6 and 7 benefit-cost ratios are also achieved well above one, indicating that the economic benefits of blocker deployment outweigh the associated costs. Scenario 6 (limited blocker deployment) produces benefit-cost ratios ranging from 16.5 to 34.4 across the four methods. The highest value again occurs under the employment-weighted Ghosh method, while the demand-side methods produce benefit-cost ratios of 16.5 and 17.0. Scenario 7 (enhanced blocker deployment) produces benefit-cost ratios ranging from 7.3 to 14.6. Although these values are lower than those observed for Scenarios 4 and 5, they still indicate strongly positive returns on investment, especially if the cheaper mitigation options have already been implemented.

## 5. Discussion

This discussion returns to the research questions to consider the broader impacts of the results reported in the previous section. Finally, a discussion of capital equipment impacts is undertaken, and direction provided for future research.

*What might be the potential economic consequences of a major space weather event in Aotearoa New Zealand?*

The results of this study provide *the first dedicated quantitative assessment* of the economic consequences of an extreme geomagnetic storm event in Aotearoa New Zealand. Across multiple scenarios, estimated GDP losses range widely depending on the geographic extent and duration of power disruption, as well as the modelling approach used. Importantly, this breadth envelopes the potential uncertainty range, making an important contribution to the literature. For example, in the absence of any mitigation measures Scenario 1 (representing a full six day national blackout) results in the highest estimated losses. Across the four modelling approaches, total GDP losses range from NZ$1.75–3.58 billion, equivalent to 0.48–0.98% of annual GDP. This range highlights the sensitivity of economic loss estimates to methodological assumptions, but consistently confirms the large-scale consequences of such a space weather event.

Even more moderate disruption scenarios, such as Scenarios 3 to 5, which simulate more geographically targeted outages informed by peak GIC exposure, still result in significant economic losses. Across these scenarios, total GDP losses range from NZ$0.65–1.48 billion, equivalent to 0.18–0.41% of annual GDP. Importantly, even under spatially limited outage assumptions, the modelling identifies substantial national-level economic consequences. This reinforces that the economic footprint of power outages extends well beyond the immediate zones of physical disruption.

Blocker deployment scenarios (6 and 7), which involve targeted hardware investments in GIC mitigation, produce the lowest economic losses. In Scenario 6, total GDP losses are reduced to NZ$0.34–0.63 billion, equivalent to 0.09–0.17% of annual GDP. In Scenario 7, losses fall further to NZ$0.24–0.48 billion, equivalent to 0.07–0.13% of annual GDP. These scenarios demonstrate the scale of economic damage that can be avoided through strategic mitigation efforts. Importantly, across all scenarios and estimation techniques, indirect losses remain a substantial portion of the total, emphasizing the need for policy and planning frameworks to move beyond a focus on direct infrastructure damages alone and account for broader system-wide interdependencies.

In sum, the economic consequences of a severe space weather event in New Zealand could reach upwards of NZ$3.58 billion in lost GDP, equivalent to 0.98% of annual GDP, in the most severe unmitigated case. Even in more localized or technically informed scenarios, economic losses remain in the range of NZ$0.65–1.48 billion, equivalent to 0.18–0.41% of annual GDP, underscoring the high economic exposure of modern electricity systems to space weather risks.

*Which mitigation investment strategies are most effective, when considering both the costs of investment, and the benefits of avoiding lost Gross Domestic Product (GDP)?*

The cost-effectiveness of different mitigation strategies can be evaluated by comparing the investment costs associated with each scenario to the corresponding reductions in economic loss across the alternative estimation approaches. Results indicate that mitigation strategies targeting early intervention and grid resilience enhancements can deliver substantial returns on investment by reducing both the duration and geographic extent of power outages.

Scenarios 4 and 5, which incorporate low-cost mitigation strategies demonstrate considerable effectiveness, such as advanced switching sequences and islanding configurations. Relative to Scenario 3, the no-mitigation GIC baseline, Scenario 4 generates benefit-cost ratios of 28.2–54.8 across the four modelling approaches, based on a one-time investment of NZ$250,000. Scenario 5 delivers the strongest overall returns, with benefit-cost ratios of 181–330 for an investment of NZ$500,000. These measures are therefore highly cost-effective, especially considering their ability to be implemented without the need for large-scale hardware changes.

Further reductions are achieved in Scenarios 6 and 7, which include physical GIC blocking devices deployed at high-risk substations and generation sites. While the associated investment costs are significantly higher, at NZ$24.75 million and NZ$68.75 million respectively, the economic benefits remain substantial. Scenario 6 generates benefit-cost ratios of 16.5–34.4 across the four modelling approaches, meaning that every NZ$1 invested avoids approximately NZ$16.5–34.4 in GDP losses. Equally, Scenario 7 produces benefit-cost ratios of 7.3–14.6, indicating that enhanced blocker deployment also remains economically justified, although the returns are lower due to the larger upfront capital cost.

Importantly, the benefits of these mitigation strategies extend beyond GDP protection. Enhanced resilience reduces the likelihood of permanent capital loss in critical infrastructure assets such as aluminium smelters and positions Aotearoa New Zealand as a proactive leader in addressing space weather hazards. When considering both direct and indirect economic losses, as well as the capital asset risks discussed in the subsequent section, the evidence clearly supports proactive investment in research, system reconfiguration (switching and islanding), and protective hardware (GIC blockers).

*Discussion of capital equipment impacts*

While the primary results presented in this study focus on direct and indirect GDP impacts from supply chain disruptions due to power outages, it is important to highlight an additional category of economic loss, which is capital damage to high-value industrial assets and associated long-term revenue losses. These represent one-off but substantial economic impacts that are not fully captured in IO modelling frameworks and VoLL approaches.

A key example is the vulnerability of large continuous industrial facilities such as aluminium smelters and steel mills, which are highly sensitive to electricity supply disruptions. NZAS at Tiwai Point presents a significant case. If electricity supply is lost for more than approximately four hours, molten aluminium inside the smelting pots would solidify, effectively destroying the production line. Restarting such an operation would require the replacement of smelter pots and related infrastructure, an undertaking estimated to cost in the region of NZ$500 million and take many months to complete. This capital loss is in addition to the foregone revenue from halting production. NZAS produces approximately 330,000 tonnes of aluminium annually, generating over NZ$1.1 billion in gross revenue. If the smelter is taken offline for a sustained period due to damage, as would be expected in a severe event without adequate mitigation, the resulting lost revenue could amount to between NZ$550 million and NZ$1.1 billion depending on the duration of the outage and recovery time, with the net economic loss likely in the range of NZ$165 million to NZ$440 million after operating costs are accounted for.

Beyond the physical replacement cost and lost production, the long-term economic output from these facilities may also be severely impacted. For example, even a temporary closure of a major smelter could lead to lost export earnings, diminished market share, employee layoffs, and downstream disruption across the regional economy. The NZAS facility directly employs around 700 staff and supports over 2,000 additional jobs through contractors and local supply chains. Closure could reduce regional GDP in Southland by an estimated NZ$100 million to NZ$200 million over a one-year period. These consequences extend well beyond the immediate blackout period and represent a structural economic loss, not just a transient shock.

Future research should specifically survey key industrial sites to estimate the replacement cost of capital equipment, installation time, and estimated lost revenues, as case study examples of strategically important facilities. In the meantime, policymakers and infrastructure operators should treat these potential capital damages as critical justifications for pre-emptive mitigation investments. Beyond this, a natural extension of this analysis would be to evaluate mitigation investments using a probabilistic net present value framework. Furthermore, it is important to also consider the impacts when faced with back-to-back space weather events during a prolonged period of activity, as this introduces challenges due to the system not recovering optimality before a new set of events. Finally, the estimates presented in this paper are only for a single system (power), therefore future assessments should look to quantify other disruptive impacts.

## Conclusions

This paper provides compelling new evidence that proactive investment in space weather mitigation can avert billions of dollars in economic losses for Aotearoa New Zealand.

In the absence of mitigation, the economic impact of an extreme but realistic geomagnetic storm could reach an upper bound of NZ$3.58 billion in lost GDP, equivalent to 0.98% of annual GDP, with approximately half of this attributable to indirect supply chain effects. Even under more moderate but spatially concentrated scenarios, losses remain substantial, ranging from NZ$0.65–1.48 billion, equivalent to 0.18–0.41% of annual GDP, highlighting the system level vulnerability of the economy to power disruptions.

Yet, relatively modest investments in mitigation can produce significant returns. Scenarios involving research-led operational strategies, such as optimized switching and islanding, deliver benefit-cost ratios as high as 330 to 1 for as little as NZ$500,000 in expenditure. Additionally, physical protections such as GIC blocking devices further reduce losses to as low as NZ$0.24–0.48 billion, equivalent to 0.07–0.13% of annual GDP, with cost-benefit returns up to 34.4 to 1. This means for every NZ$1 invested in targeted physical mitigation, up to NZ$34.4 in GDP losses can be avoided, while operational mitigation can avoid up to NZ$330 for every NZ$1 invested. When also considering the unmodelled impacts on capital intensive assets like the NZAS, where combined capital and revenue losses could exceed NZ$1 billion, the case for early, targeted mitigation becomes even more urgent.

These results make a best estimate of the potential impacts under the proposed scenarios, but it is important to state that any outcomes are dependent on a range of uncertain factors, including CME-geospace interactions, ground resistivity, network configuration and operation, and human behavioural responses to power outages. Importantly, we contribute to the literature a range of scenarios which indicate the economic envelope for an extreme, but plausible and realistic, space weather event.

These findings make a strong economic argument for embedding space weather resilience into infrastructure planning, and they underscore the broader policy imperative to treat geomagnetic storm preparedness as a national economic priority. Future research needs to focus on modelling capital and revenue losses for strategically important economic facilities.

Acknowledgements

EJO and DB are supported by funding from NSF RAPID grant (#2434136) and the NSF National Center for Atmospheric Research (#1852977) via the Early-Career Faculty Innovator Program. DHM and CJR were supported by the New Zealand Ministry of Business, Innovation and Employment Endeavour Fund Research Programme contract UOOX2002. The authors would like to thank Transpower New Zealand for supporting this study by the provision of data, in-kind labour time, and modest salary support. The authors thank Terry Griffin of Kansas State University for informally reviewing an early version of the working paper.

Data availability statement

All code for the economic impact assessment can be found at the New Zealand Space Weather Assessment (NZ-SWA) GitHub repository (https://github.com/edwardoughton/nz-swa) and the associated Zenodo repository for processed data and results (https://zenodo.org/records/18071860). The New Zealand electrical transmission network's characteristics were provided to the University of Otago team by Transpower New Zealand with caveats and restrictions. This includes requirements of permission before all publications and presentations, and no ability to provide the observations themselves. In addition, we are unable to provide the New Zealand network characteristics due to commercial sensitivity. Requests for access to these characteristics need to be made to Transpower New Zealand. At this time the contact point is Michael Dalzell (Michael.Dalzell@transpower.co.nz).

Supplementary information

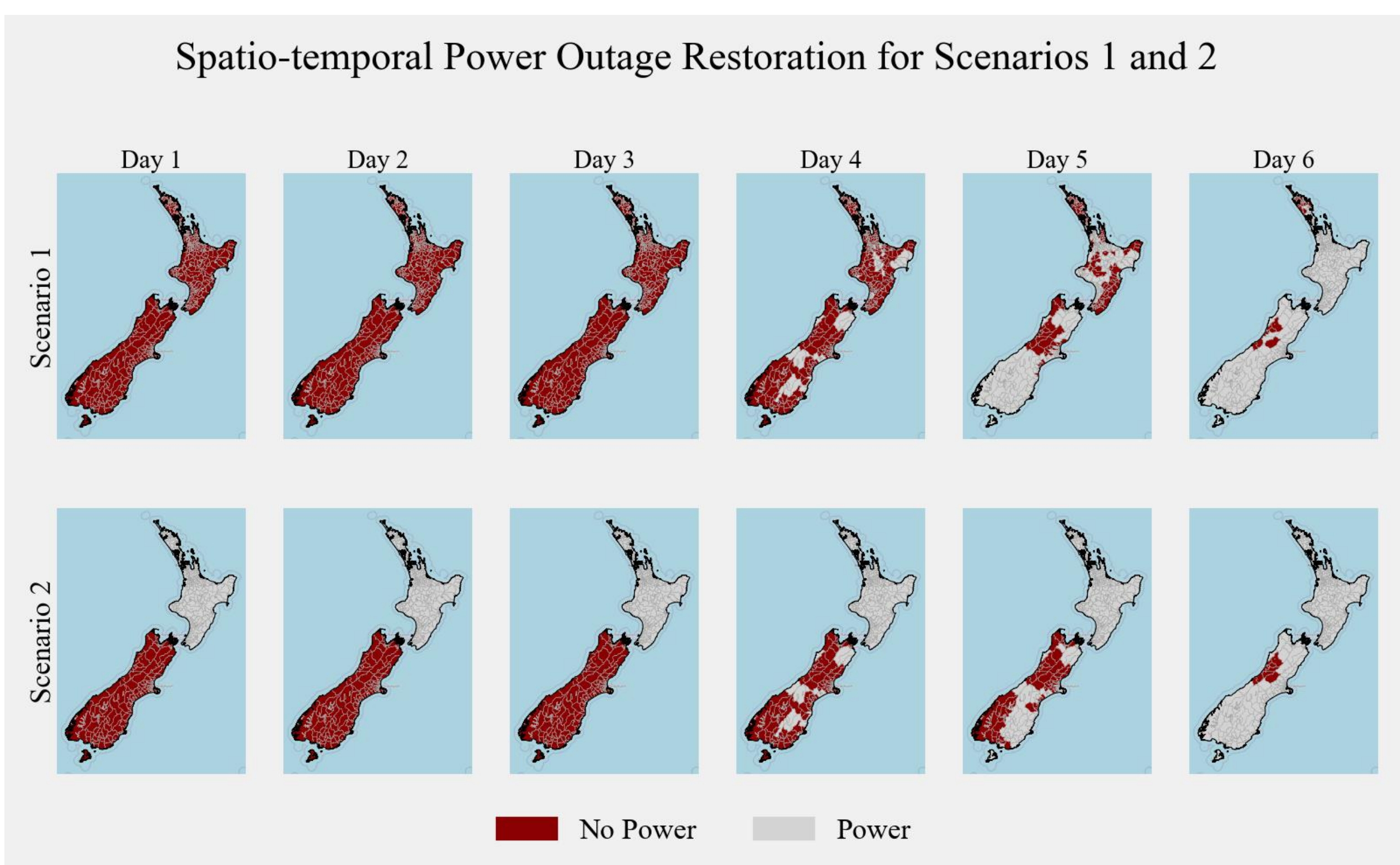

Figure S1 Restoration sequence for scenarios 1 and 2

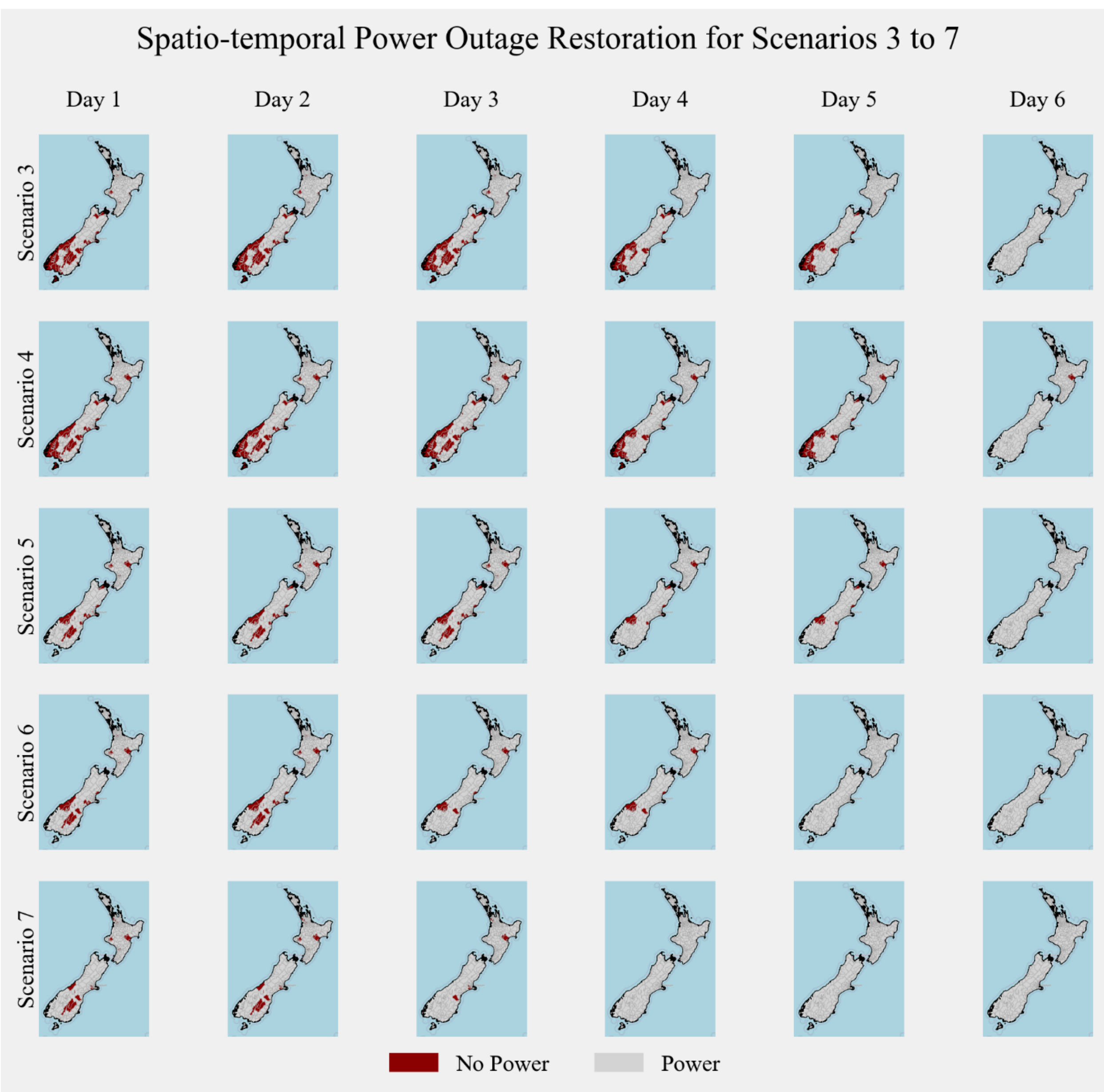

Figure S2 Restoration sequence for scenarios 3-7

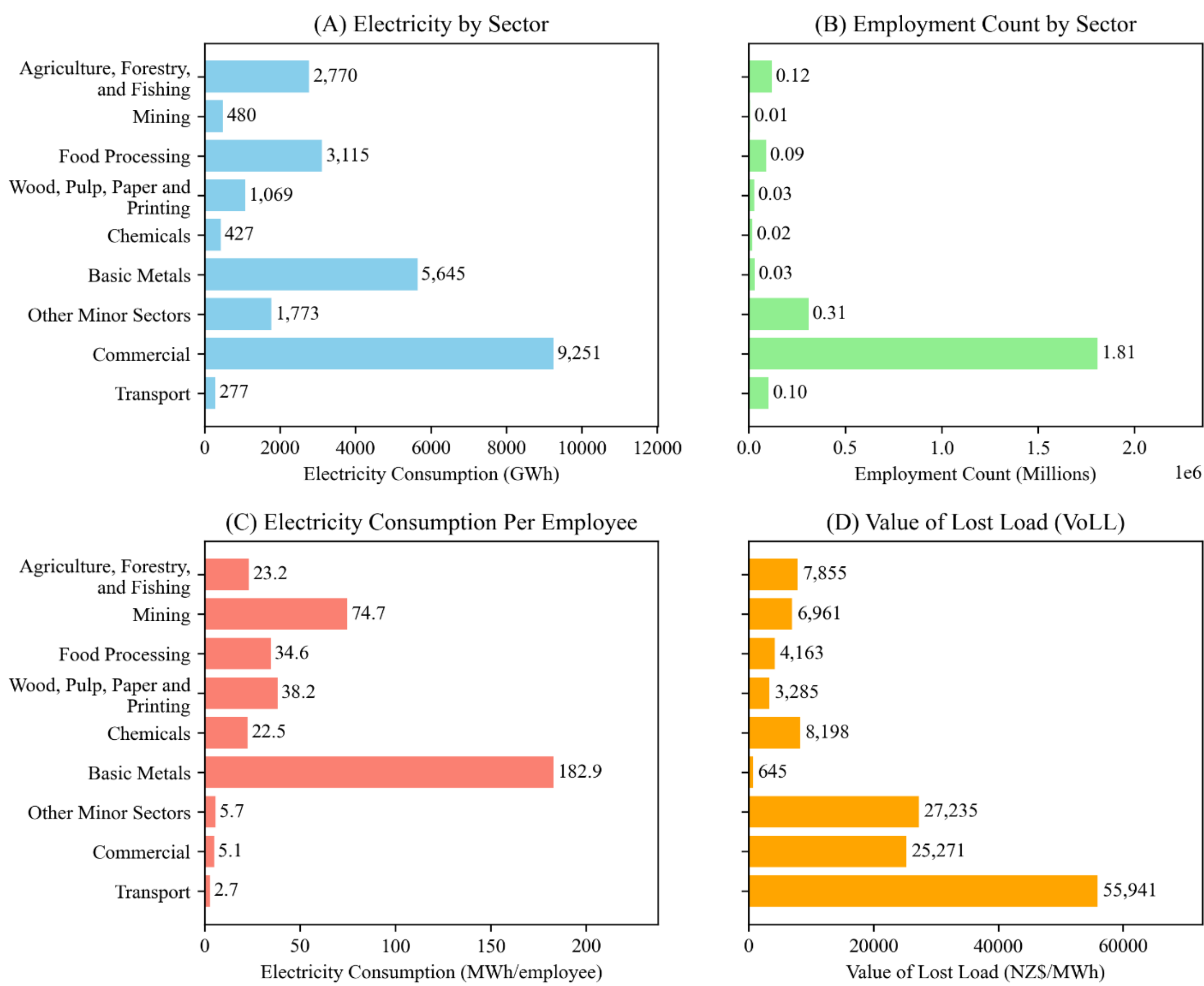

Figure S3 Electricity consumption and the VoLL in New Zealand

Lost Direct and Indirect GDP by Industrial Sector (Demand-Side Leontief, Population Shock)

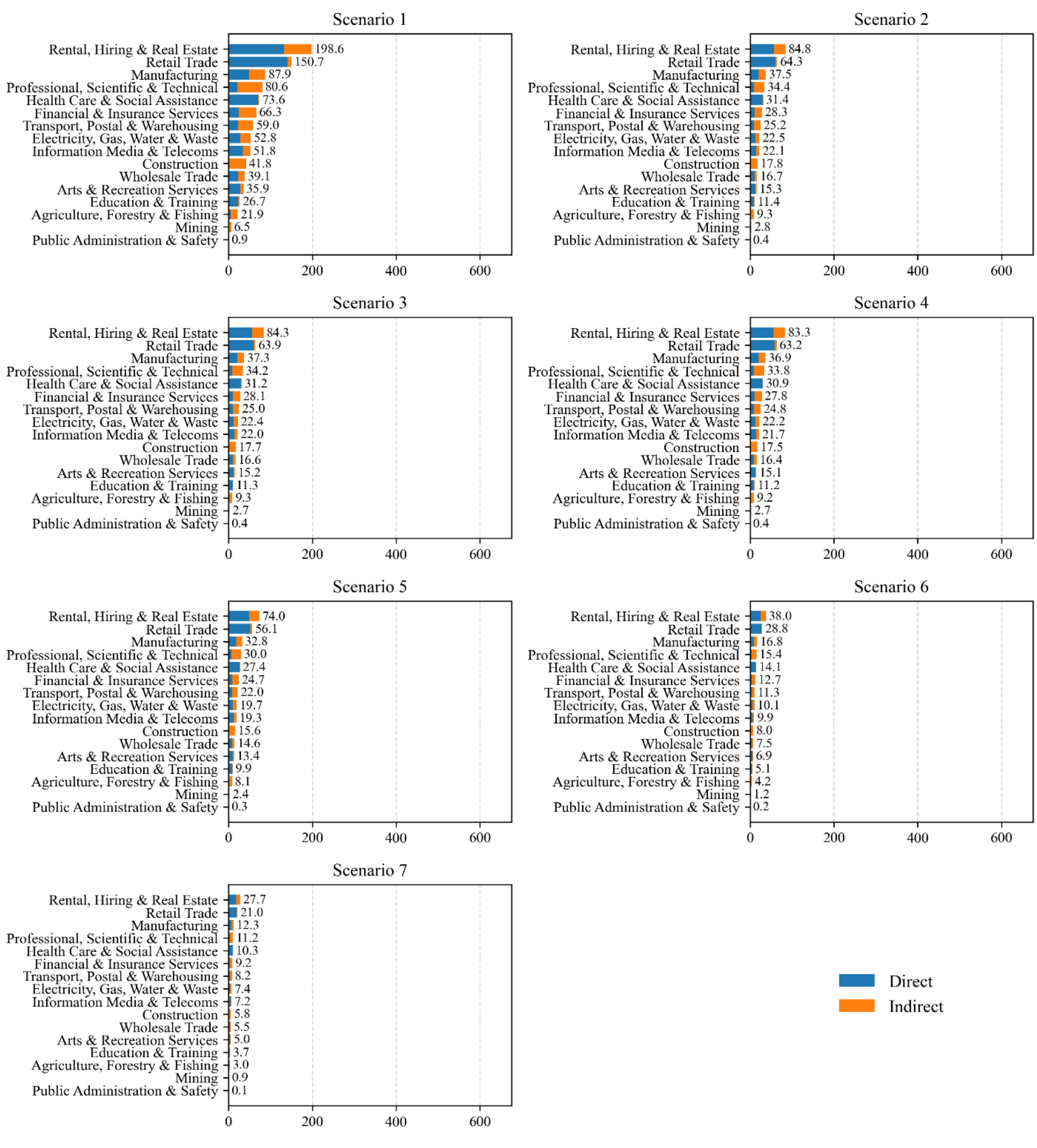

S4 Sector GDP impacts using a Demand-Side Leontief population shock

Lost Direct and Indirect GDP by Industrial Sector (Demand-Side Leontief, Survey-Based VoLL)

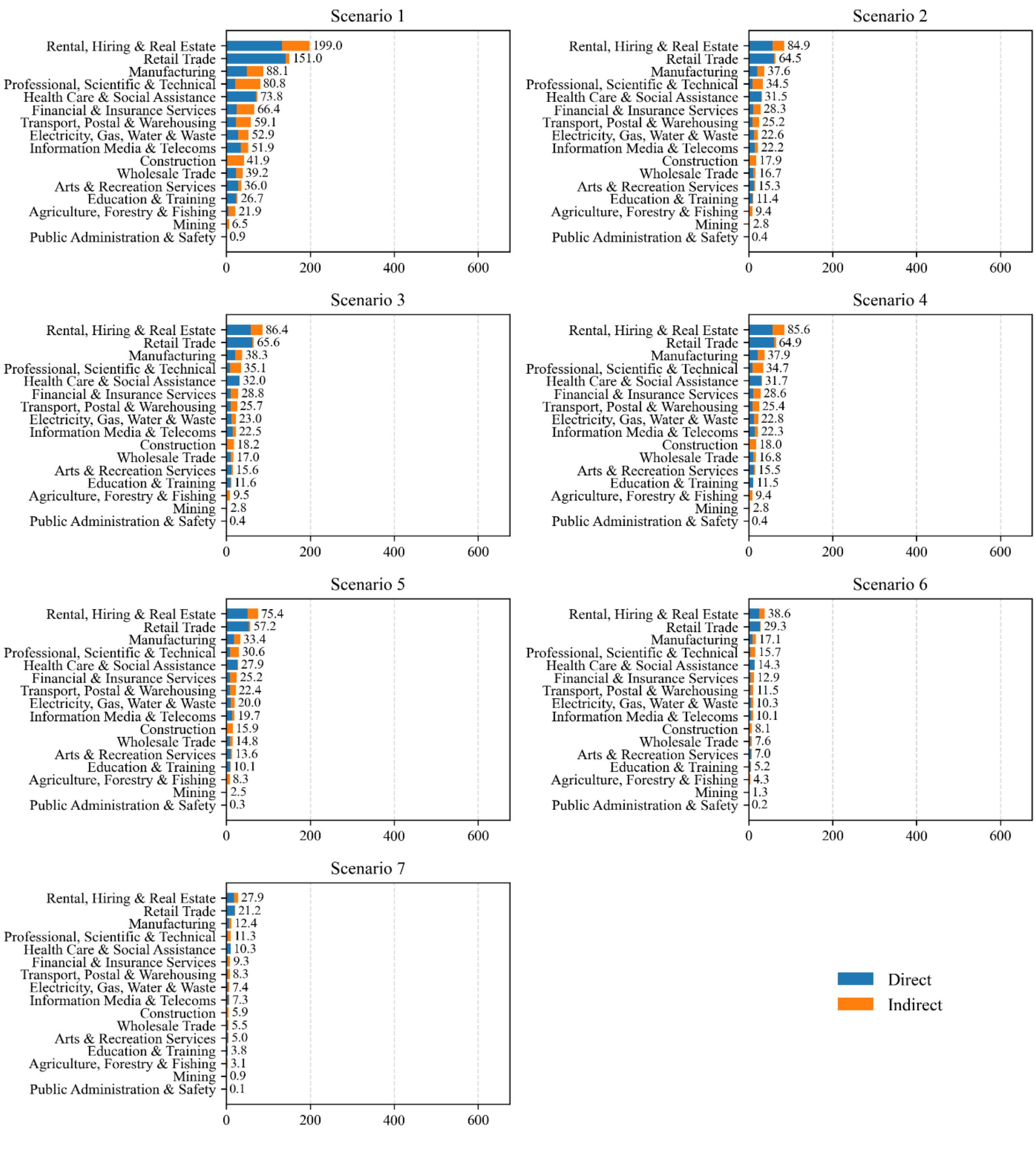

<u>S5 Sector GDP impacts using a Demand-Side Leontief survey-based VoLL shock</u>

Lost Direct and Indirect GDP by Industrial Sector and Scenario (Supply-Side Ghosh, Employment Shock)

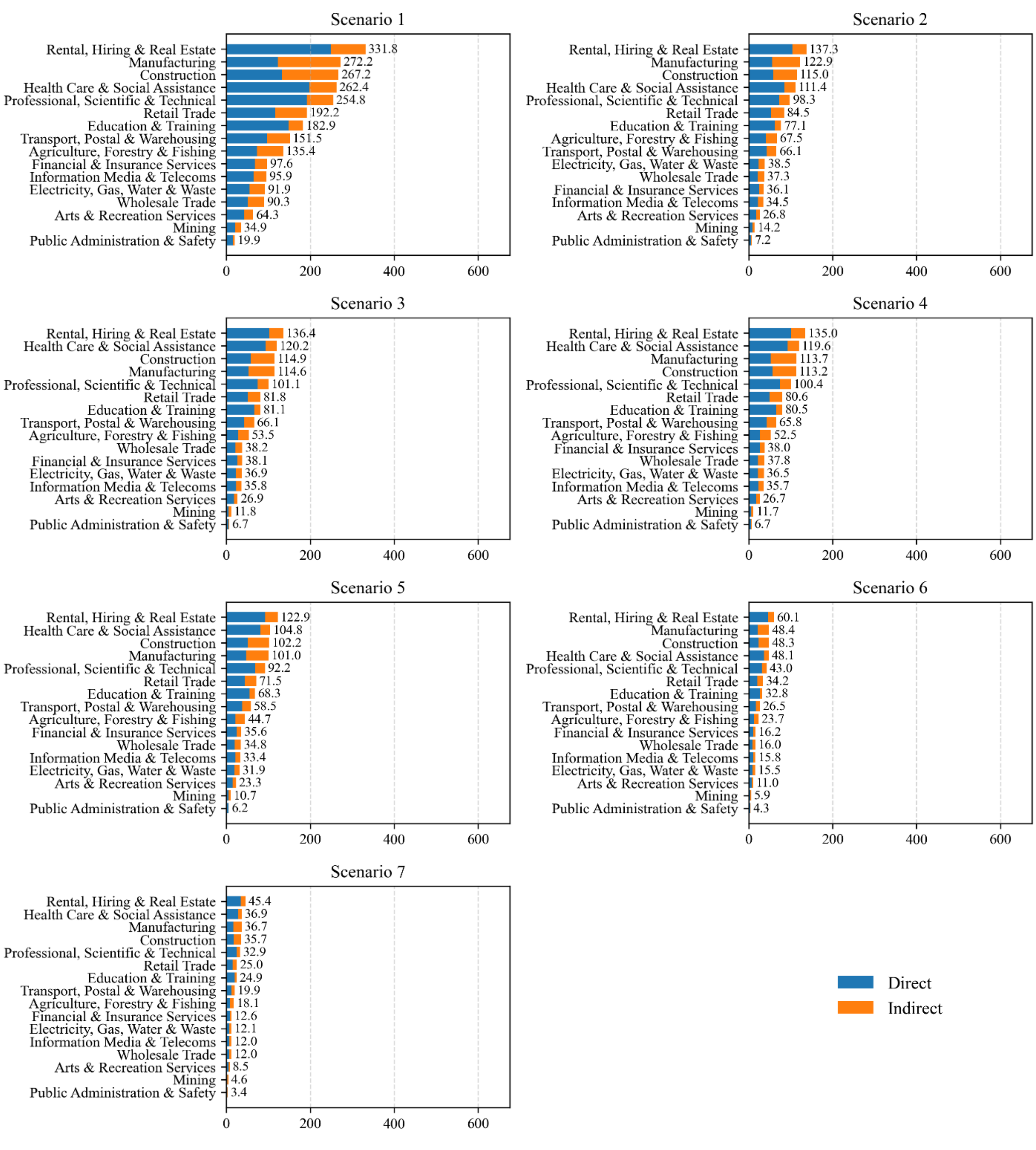

Lost GDP (Millions 2026 NZ$)

<u>S6 Sector GDP impacts using a Supply-Side Ghosh employment shock</u>

Lost Direct and Indirect GDP by Industrial Sector and Scenario (Supply-Side Ghosh, Survey-Based VoLL)

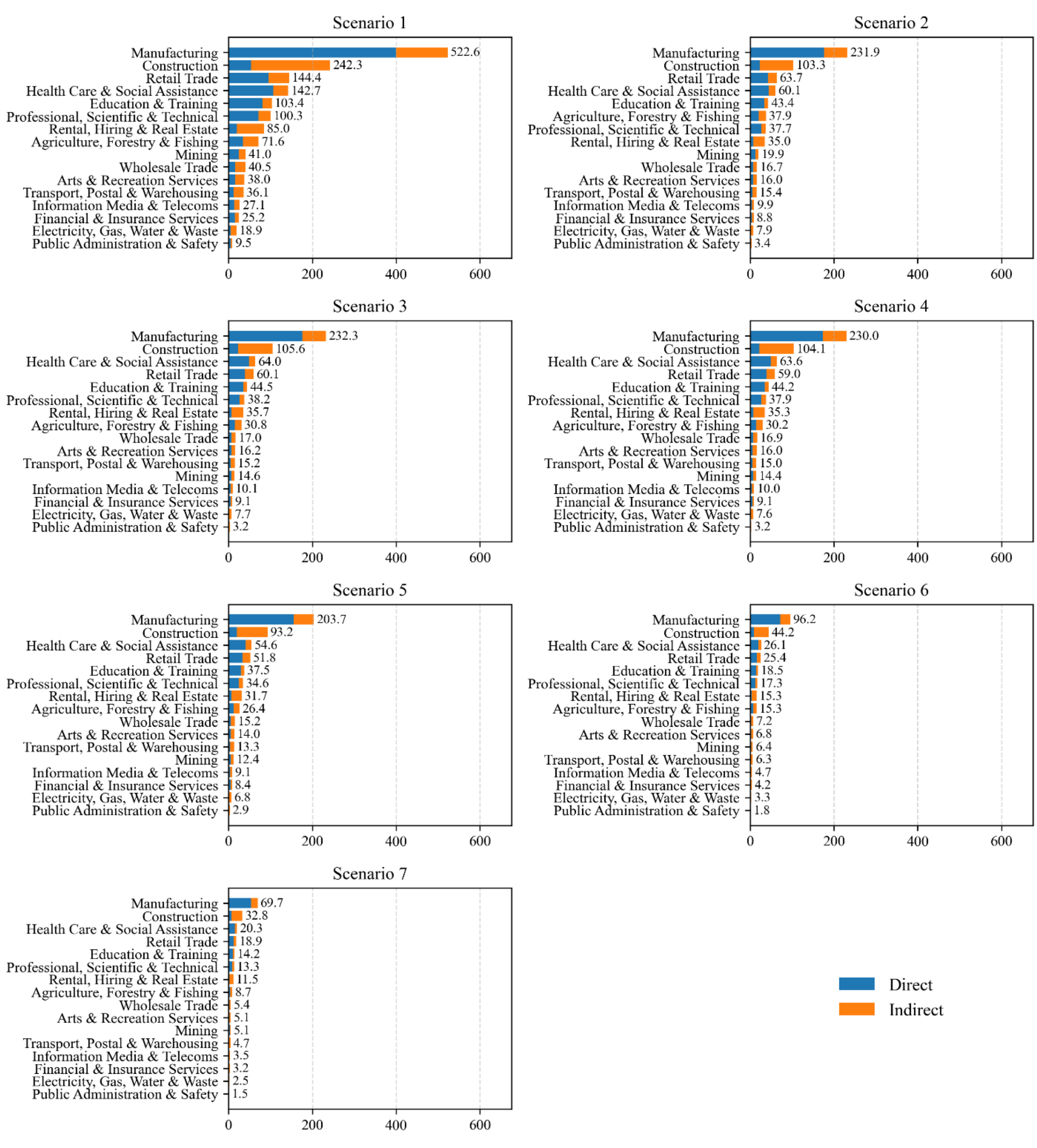

Lost GDP (Millions 2026 NZ$)

<u>S7 Sector GDP impacts using a Supply-Side Ghosh survey-based VoLL</u>